\documentclass[10pt,letterpaper]{article}

\usepackage[top=0.85in,left=2.75in,footskip=0.75in]{geometry}

\usepackage{amsmath,amssymb}

\usepackage{changepage}

\usepackage[utf8x]{inputenc}

\usepackage{textcomp,marvosym}

\usepackage{cite}

\usepackage{nameref,hyperref}

\usepackage[right]{lineno}

\usepackage{microtype}
\DisableLigatures[f]{encoding = *, family = * }

\usepackage[table]{xcolor}

\usepackage{array}

\newcolumntype{+}{!{\vrule width 2pt}}

\newlength\savedwidth

\raggedright
\usepackage[aboveskip=1pt,labelfont=bf,labelsep=period,justification=raggedright,singlelinecheck=off]{caption}

\makeatletter
\renewcommand{\@biblabel}[1]{\quad#1.}
\makeatother

\usepackage{lastpage,fancyhdr,graphicx}
\usepackage{epstopdf}
\fancyheadoffset[L]{2.25in}
\fancyfootoffset[L]{2.25in}
\begin{document}
\vspace*{0.2in}

\begin{flushleft}
{\Large
\textbf\newline{Determining the number of factors in a forecast model by a random matrix test: cryptocurrencies} 
}
\newline
\\
Andrés García-Medina\textsuperscript{1,2*},
Graciela González-Farías\textsuperscript{2},
\\
\bigskip
\textbf{1} Consejo Nacional de Ciencia y Tecnología, Av. Insurgentes Sur 1582, Col. Crédito Constructor 03940, Ciudad de México, México
\\
\textbf{2} Probabiliy and Statistics, Centro de Investigación en Matemáticas, A.C. Jalisco S/N, Col. Valenciana 36240, Guanajuato, Mexico
\\

\bigskip

* andres.garcia@cimat.mx

\end{flushleft}
\section*{Abstract}
We determine the number of statistically significant factors in a forecast model using a random matrices test. The applied forecast model is of the type of Reduced Rank Regression~(RRR), in particular, we chose a flavor which can be seen as the Canonical Correlation Analysis~(CCA). As empirical data, we use cryptocurrencies at hour frequency, where the variable selection was made by a criterion from information theory. The results are consistent with the usual visual inspection, with the advantage that the subjective element is avoided. Furthermore, the computational cost is minimal compared to the cross-validation approach.

\section*{Introduction}
Cryptocurrencies are new financial instruments which are based on the technology of
blockchains~\cite{Nakamoto}. 
A coin is defined as a chain of digital signatures. Each owner transfers the coin to the next by digitally signing a hash of the previous transaction and the public key of the next owner and adding these to the end of the coin. The easy access of this new financial instrument through more than 17000 exchanges with low fees of transactions, more than 2000 virtual currencies worldwide and a  traded volume of nearly 60 billion dollars, have done cryptocurrencies a very attractive instrument of investment for the general population~\cite{coinmarketcap}.

There have been previous attempts to characterize the collective behavior of cryptocurrencies as is the work~\cite{Stosic2018}. There it is shown that a large data set of cryptocurrencies at daily frequency deviate from the universal results of Marchenko-Pastur~\cite{Marchenko}. In addition, the study state that the spanning tree structure is stable over time. Further, in the work~\cite{Stanley2018} is analyzed the power-law behavior of Bitcoin for a large period of time and different frequency levels, from one minute to one day. They conclude that Bitcoin exhibit heavy-tails in the range $2<\alpha <2.5$ across multiple coin exchanges. Their findings support the use of standard financial because of the finite variance implications of the results.

On the contrary, the aim of this work is to provide tools related to the forecast and invest problems by combine mathematical tools apparently unrelated, and having as a data sample the new cryptocurrency instruments.
Thus, the proposed methodology is general and can be applied to any data set for which there is interest to analyze.

In the next section the preprocessing of the data set of cryptocurrencies is presented.
Next, in the section called variable selection is proposed the use of the transfer entropy measure from information theory  to discriminate between  the set of predictor and response variables, i.e. to solve the variable selection problem which is inherent to any forecast model.
In the forecast model section is introduced the general regression model where the studied model is framed.
Afterward, random matrix theory is used to select the proper number of factors in the presented multi-response regression model when working at the high dimensional level. Then, in the  number of factors section is described the mathematical relation of some results in high dimensional statistics with the reduced rank selection problem for the particular case of canonical correlation analysis.
Finally, in the conclusion section the main findings are summarized and future work is proposed.

\section*{Data}

A sample of $p = 100$ cryptocurrencies is taken using the API of CoinMarketCap~\cite{coinmarketcap}, on the elapsed period from  May 23 to November 27 of 2018 at frequency of hours, given a total of $n = 4533$ observations~(see \nameref{S1_File} and \nameref{S1_Table}). 
We work with returns of the standardized prices $Z_k(t)$ for every cryptocurrency~($k=1,\dots, p$) and time~($t=1,\dots,n$)
  \begin{equation}
    R_k(t) = \frac{Z_k(t+\Delta t) -Z_k(t)}{Z_k(t)},
  \end{equation}
In this manner, the Augmented Dickey-Fuller test~\cite{Dickey} assures that the involved time series are stationary with a \textit{p-value} less than $0.01$ for all the return times series $R_k$~($k=1,\dots,p$) considered.

\section*{Variable selection}

One of the first problems when trying to set a forecast model is the variable selection problem.
Usually, in the econometric approach, the economic theory dictates which variables must be treated as a predictors and which as a response.
However, cryptocurrencies are a new financial instrument for which there are not many economic models behind them.
Hence, we follow an information approach to solve the variable selection problem.

In 2000 T. Schreiber introduced the quantity Transfer Entropy~(TE) in the context of information theory with the purpose of measuring the information flow from one process to another in a non symmetrical way.
Let $x_i = x(i)$ and $y_i = y(i), i=1,\dots, N$, denote sequences of observations from
systems $X$ and $Y$. TE is defined as~\cite{Schreiber}
\begin{equation}
  T_{Y\rightarrow X}(k,l) = \sum_{i,j}  p(x_{t+1}, x_t^{(k)}, y_t^{(l)} ) \log \frac{p(x_{t+1} | x_t^{(k)}, y_t^{(l)})}{p(x_{t+1} | x_t^{(k)})},
\end{equation}

The idea behind TE is to incorporate time dependence by relating previous samples $x_i$ and $y_i$ to predict
the next value $x_{i+1}$, and quantify the deviation from the generalized Markov property, $p(x_{i+1}|x_i,y_i) = p(x_{i+1}|x_i)$,
where $p$ denotes the transition probability density. 
If there is no deviation from the generalized Markov property, $Y$ has no influence on $X$.
TE, which is formulated as the Kullback-Leibler entropy~\cite{Kullback} between $p(x_{i+1}|x_i,y_i)$ and $p(x_{i+1}|x_i)$ quantifies the incorrectness of this assumption, and is explicitly nonsymmetric under the exchange of $x_i$ and $y_i$.

An interesting property of TE is that under some conditions it can be seen as a non-linear generalization of Granger causality. 
In econometrics, Granger causality plays an important role in the parameter estimation of a vector autoregressive~(VAR) model.
Granger causality has as an assumption that cause precedes effect, and a cause have information about the effect that is unique and no present on other variable.

Consider the jointly stationary stochastic processes $X_t, Y_t$.
Let $F\left(x_t|x_{t-1}^{(k)}, y_{t-1}^{(l)} \right)$ denote the distribution function of the target variable $X$ conditional on the joint~($k$,$l$)-history $X_{t-1}^{(k)}, Y_{t-1}^{(l)}$.
Then, variable $Y$ is said to Granger-cause variable $X$~(with lags $k,l$) if and only if~\cite{Granger, TE_Book}
  \begin{equation}
   F\left(x_t|x_{t-1}^{(k)}, y_{t-1}^{(l)} \right) \neq F\left(x_t|x_{t-1}^{(k)} \right).
  \end{equation}
Thereby, it is said that $Y$ Granger-causes $X$ if and only if $X$ is not independent of the history of $Y$.

There exist a series of results~\cite{Barnett2009, Schindler2011, Barnett2012} which state an exact equivalence between the Granger causality and TE statistics for different approaches and assumptions of the data generating processes, which enable to construct TE as a non-parametric test for pure Granger causality.
This connection can be seen as a bridge between causal inference of data under autoregressive models and the information theory approach.
Before proceed we want to emphasize that for highly non-linear and non-Gaussian data as is the case of many financial instruments, it is better to approach causality by TE information method instead of the traditional Granger causality test~\cite{TE_Book}.

In real data applications we need to estimate TE from observed data.
There are several techniques to estimate TE from observed data, however most of them make a great demand on the data Nevertheless and consequently are commonly biased due to small sample effects, which limit the use of TE to real data applications.
To avoid this problem, we use the robust and computationally fast technique of symbolization~\cite{Staniek} to estimate TE.
Symbolic Transfer Entropy~(STE) has been introduced within the concept of permutation entropy~\cite{Bandt}.

Following~\cite{Bandt, Staniek}, symbols are defined by reordering the amplitude values of time series $x_i$ and $y_i$ .
Thus, for a given $i, m$  arbitrary  amplitude values, the elements 
 \begin{equation}
       \{x(i), x(i+l), \dots, x(i+(m-1)l)\},
 \end{equation}
are arranged in an ascending order
  \begin{equation}
   \{x(i+(k_{i1}-1)l)\leq x(i+(k_{i2}-1)l)\leq \dots \leq x(i+(k_{im}-1)l)\},
  \end{equation}
where $l$ denotes the time delay, and $m$ the embedding dimension.
A symbol is thus defined as $\hat{x}_i = (k_{i1}, k_{i2},\dots,k_{im})$, and with the relative frequency of symbols is estimated the joint and conditional probabilities of the sequence of permutation indices.

To exemplify this procedure let us take the  time series $\{1, 2, 3, 6, 5, 4\}$ to estimate the related Shannon entropy measure of information theory~\cite{Shannon}.
First, we need to organize the five pairs of neighbors according to their relative values. 
Thereby, it is found three pairs which satisfies the relation $x_t <  x_{t+1}$ characterized by the permutation $\{01\}$, and two pairs for which $x_t > x_{t+1}$ represent the permutation $\{10\}$.
Then, the Shannon entropy for $m=2$ is given by
\begin{equation}
 H(2) = -(3/5)\log(3/5) -(2/5)\log(2/5) \approx 0.971.
\end{equation}

Let us now go back to the original problem of TE estimation.
Given symbol sequences $\{\hat{x}_i\}$ and $\{\hat{y}_i\}$,  STE is mathematically defined as\cite{Staniek}
  \begin{equation}
   T^{S}_{Y\rightarrow X} = \sum_{i,j}  p(\hat{x}_{i+\delta}, \hat{x}_i, \hat{y}_i ) \log \frac{p(\hat{x}_{i+\delta} | \hat{x}_i, \hat{y}_i)}{p(\hat{x}_{i+\delta} | \hat{x}_i)},
  \end{equation}
where the sum runs over all symbols and $\delta$ denotes a time step. The log is with base 2, thus $T^{S}_{Y\rightarrow X}$ is given in bits.

The question at this point is whether a given empirical measurement of STE is statistically different from 0, and represents sufficient evidence for a  direct relationship between the variables.
It is possible to construct a null hypothesis $H_0$ that there is no such relationship, but is necessary to know what the distribution for the empirical measurement would look like if $H_0$ were true, and then evaluate a \textit{p-value} for sampling the actual measurement from the distribution.
If the test fails, we accept the alternate hypothesis that there exists a directed relationship.

For discrete $X$ and $Y$, it is know that if $H_0$ is true then $T^{S}_{Y^s \rightarrow X} \xrightarrow{d} \chi^2(D)/(2N \log 2)$, where the number of degrees of freedom $D$ is the difference between the number of
parameters in the full and null models~\cite{Barnett2012}.
$Y_s$ represents surrogate variables for $Y$ generated under $H_0$, which have the same statistical properties as $Y$ , but any potential correlation with $X$ is destroyed.
As a consequence, surrogates of the distribution  $T^{S}_{Y^s \rightarrow X}$ must preserve $p(\hat{x}_{i+\delta} | \hat{x}_i)$ but not $p(\hat{x}_{i+\delta} | \hat{x}_i, \hat{y}_i)$~\cite{JIDT}.

In order to present our results in the context of a forecast model, let us rename the variable $x(t)$ as the predictor and the variable $y(t)$ as the response. 
Thus, we estimate STE  for the combination of pairs $\{X_{a}(t), Y_{b}(t+\Delta t)\}$, where $a,b = 1,\dots, p~(=100)$; $t = 0,\dots, n-\Delta t$, being $\Delta t$ an added lag time to consider forecast situations. The results for a time delay $l=1$ and \textit{p-value} $=0.10$ are given in Table~\ref{table1} for different values of lag time $\Delta t$ and embedding dimension $m$. In the third column it is shown the total sum of $T^{S}_{X^s_a \rightarrow Y_b}$ for all the possible combinations of the indices $a,b$ as long as exist a direct relationship under $H_0$. In the four column it is shown the number of relations which are preserved. We found a peak in the number of preserved relations at $\Delta t = 1$ and $m=2,3$, having more than 7000 relations out of the 10000 possible relations. Even though the maximum is reached at $m=2$, we chose the case $m=3$ following the criterion of get at the same time the maximum of the total sum of information flow~(118.1084 bits).

\begin{table}[!ht]
\centering
\caption{
{\bf STE results}}
\begin{tabular}{|l|l|l|l|}
\hline
\textbf{$\Delta t$} & \textbf{$m$} & \textbf{$\sum_{ab} T^{S}_{ab}$} & \textbf{\#\{$T^{S}_{ab}>0$\}} \\ \hline
0            & 2          & 7.7484          & 4221                            \\ \hline
0            & 3          & 97.7024         & 6345                            \\ \hline
0            & 4          & 241.6957        & 736                             \\ \hline
\textbf{1}    & \textbf{2}  & \textbf{19.677}  & \textbf{7756}                            \\ \hline
\textbf{1}   & \textbf{3} & \textbf{118.1083}& \textbf{7067}                            \\ \hline
1            & 4          & 351.52          & 1069                            \\ \hline
2            & 2          & 1.3937          & 1289                            \\ \hline
2            & 3          & 68.196          & 4701                            \\ \hline
2            & 4          & 442.0707        & 1342                            \\ \hline
3            & 2          & 1.3346          & 1240                            \\ \hline
3            & 3          & 13.8508         & 1070                            \\ \hline
3            & 4          & 333.1614        & 1013                            \\ \hline
\end{tabular}  
\begin{flushleft} STE of cryptocurrency return time series for lag times $\Delta t = 0,1,2,3$ of the pair predictor-response variables $X$,$Y$ respectively; and embedding dimension $m=2,3,4$. In the third column it is shown the total amount of direct information  at the \textit{p-value} of $0.10$, while in the fourth column  it is shown the corresponding number of preserved relations at the same level of statistical significance.
\end{flushleft}
\label{table1}
\end{table}

\begin{figure}[!h]
\includegraphics[height=0.3\textheight]{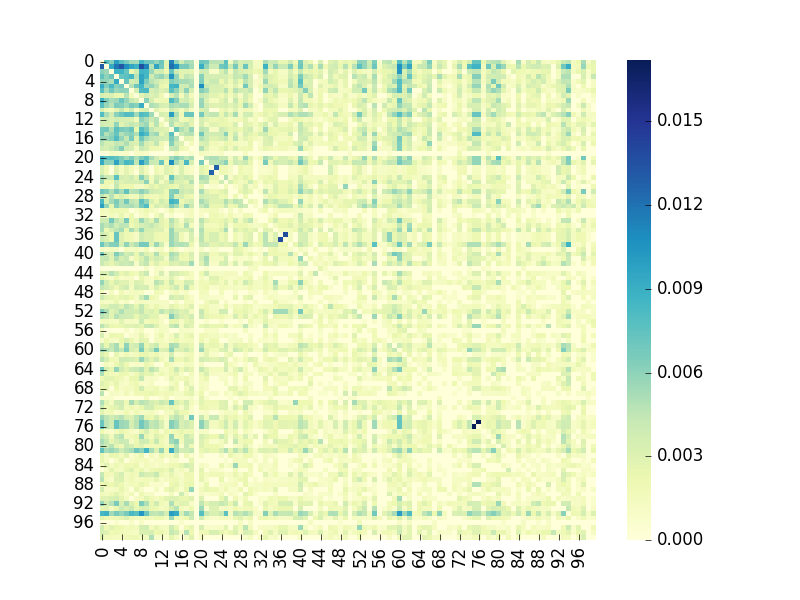}
\caption{{\bf Heatmap of STE for $m=2$.}
The color intensity represent the magnitude of STE.}
\label{fig1}
\end{figure}

\begin{figure}[!h]
\includegraphics[height=0.3\textheight]{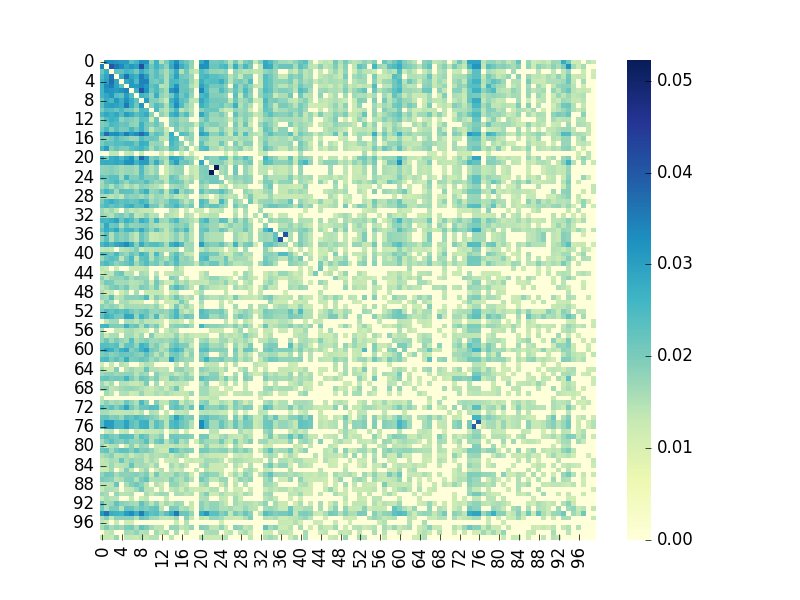} 
\caption{{\bf Heatmap STE for $m=3$.}
The color intensity represent the magnitude of STE.}
\label{fig2}
\end{figure}

Moreover, we show in Fig~\ref{fig1} and Fig~\ref{fig2} the heatmap of STE results for $m=2$ and $m=3$, respectively.
It can be appreciated higher values of STE in Fig~\ref{fig2} than in Fig~\ref{fig1} in general.
Further, it can be noticed some structure in the upper left of Fig~\ref{fig1}, which is sharper in Fig~\ref{fig2}.
This upper left section refers to the cryptocurrencies with the highest capitalization due to the way we order them. 
Therefore, it is natural to have the highest values of information flow in that sector.

A convenient procedure to measure the net flow of information between the processes $X$ and $Y$ is by the normalized directionality index~(NDI), given by~\cite{TE_Book}
 \begin{equation}
  d(X,Y) = \frac{STE_{X \rightarrow Y} - STE_{Y\rightarrow X}}{STE_{X \rightarrow Y} + STE_{Y\rightarrow X}}\in[-1,1]
 \end{equation}

\begin{figure}[!h]
\includegraphics[height=0.4\textheight]{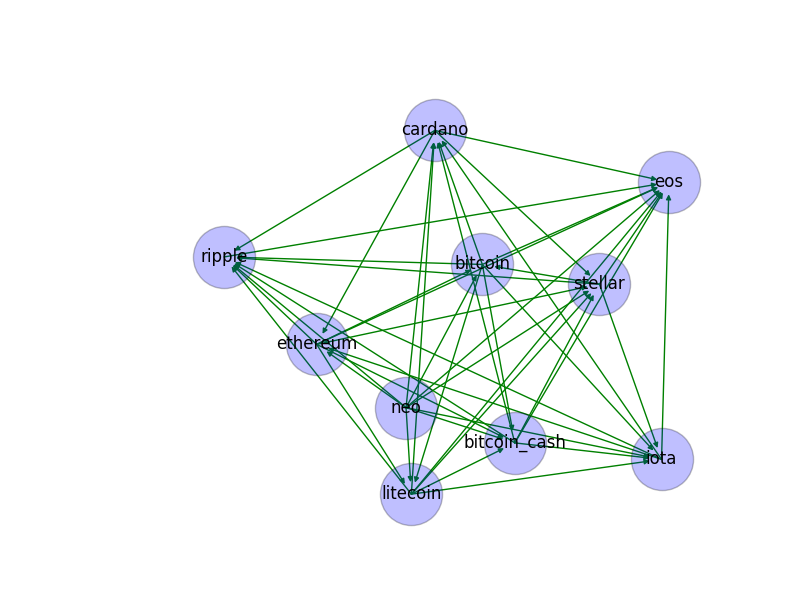}  
\caption{{\bf  NDI subgraph.}
The arrow direction represents the direction of the information flow.}
\label{fig3}
\end{figure}

This quantity regularizes $STE$ values between $-1$ and $1$, such that $d(X,Y)$ is maximized when one of the $STE$ values is zero and minimized when are equal.
This index is not normalized in the statistical sense, but it resembles a measure of divergence or market leverage, and beyond that, it is very useful to compare measures across different systems or financial sectors.
We applied NDI to our previous results for $\Delta t = 1$ and $m=3$.
In order to have a better visualization, the obtained STE values are first converted to a directed graph $G=(V,E)$, where the  nodes $V$ are the different cryptocurrencies and the edges $E$ the resulting value $d(X,Y)$ of applying NDI.
In Fig~\ref{fig3} it is shown as an example a directed subgraph with the first 10 cryptocurrencies in capitalization order with its corresponding edges given by the measure NDI. There, the arrow direction tells us how the information flows from one variable to another and as a consequence more dominant. We can see for example that the coin \emph{eos} only receive information from the other coins under the measure NDI, whereas \emph{ripple} send and receive information from the members of the subnet.

In order to discriminate the predictor variables from the response variables, some basic concepts of graphic theory were used.
The node out-degree is the number of edges pointing out the node, while the node in-degree is the number of edges pointing into the node.
We used these concepts to select the sets of predictor-response variables by the proposed heuristic selection rule:
   \begin{itemize}
    \item  $V_i \in \{\text{response variables}\}$ if \#in-degree  $\geq$ \#out-degree,
    \item  $V_i \in \{\text{predictor variables}\}$ if \#in-degree $<$ \#node out-degree,
   \end{itemize}
for $i=1,\dots, p$. The results of applying this procedure are shown in Table~\ref{table2} for the first 10 response and predictor variables~(see \nameref{S2_Table} for the entire list). In general, we found 49 predictor variables and 51 response variables in our set of $p=100$ return times series of cryptocurrencies.

\begin{table}[!ht]
\centering
\caption{
{\bf Predictor and response variables.}}
\begin{tabular}{|l|l|l|}
\hline
\textbf{i} & \textbf{Predictor~(49)} & \textbf{response~(51)} \\ \hline
1          & ethereum           & bitcoin            \\ \hline
2          & neo                & ripple             \\ \hline
3          & dash               & bitcoin cash      \\ \hline
4          & monero             & litecoin           \\ \hline
5          & lisk               & cardano            \\ \hline
6          & bitcoin gold       & stellar            \\ \hline
7          & tether             & eos                \\ \hline
8          & steem              & iota               \\ \hline
9          & populous           & nem                \\ \hline
10         & siacoin            & ethereum classic  \\ \hline
\vdots     & \vdots   	        & \vdots             \\ \hline
\end{tabular}
\begin{flushleft} First 10 predictor and response variables in capitalization order, which are selected under the heuristic criterion given above. The total number of selected variables is shown in parentheses.
\end{flushleft}
\label{table2}
\end{table}

Now, once found the set of predictor-response variables, we would like to present the general regression model which has been used as the framework to forecast our response variables. Thus, in the next section is presented this model and the related problem of rank determination which bring up the necessity to study some results of random matrices.

\section*{Forecast model}

Consider the Reduced Rank Regression~(RRR) model given by~\cite{Izenman1975}
\begin{equation}
 \overset{s\times 1}{\mathbf{Y}} = \overset{s\times 1}{\mathbf{\mu}} + \overset{s\times r,}{\mathbf{C}} \space\space \overset{r\times 1}{\mathbf{X}} + \overset{s\times 1}{\mathbf{\varepsilon}}
\end{equation}
where $\mathbf{\mu}$ and $\mathbf{C}$ are unknown regression parameters, the unobservable error variate $\varepsilon$ of the model has mean $E(\varepsilon) = 0$, covariance matrix $cov(\varepsilon) = E\{\varepsilon\varepsilon^{\tau}\} = \Sigma_{\varepsilon\varepsilon}$, and is distributed independently of $\mathbf{X}$.

The difference with the classical multivariate regression model is that the rank of the regression coefficient matrix $\mathbf{C}$ is deficient
\begin{equation}
 \text{rank}(\mathbf{C}) = t \leq \text{min}(r,s).
\end{equation}
The rank condition implies that there may be a number of linear constraints on the set of regression coefficients in the model.

Given a sample $\mathbf{X,Y}$ of observations, the goal is to estimate the parameters $\mathbf{\mu}$ and  $\mathbf{C}$ in an optimal manner.
Hence, the idea is to minimize the objective function
  \begin{equation}
   W(t) = E\{(\mathbf{Y - \mu - CX})^{'}\mathbf{\Gamma(Y-\mu - CX)}\},
  \end{equation}
where $\mathbf{\Gamma}$ is a positive-definitive symmetric matrix of weights and the
expectation is taken over the joint distribution of $\mathbf{X,Y}$.

RRR can be seen as a unifying treatment of several classical multivariate procedures that were developed separately from each other.
If we set $\mathbf{X} \def \mathbf{Y}$ (and $r = s$) by making the output variables identical to the input variables, and in addition set $\mathbf{\Gamma} = \mathbf{I}$, then we have Harold Hotelling’s
principal component analysis and exploratory factor analysis.
If we set $\mathbf{\Gamma = \Sigma_{YY}^{-1}}$, then we have Hotelling’s canonical variate and correlation analysis.
A nonlinear generalization of RRR provides a flexible model for artificial neural networks~\cite{IzenmanBook}.

Nevertheless, one of the primary and most difficult parts of the model determination is  to assess the unknown value of the parameter $t$, which is called the effective dimensionality of the multivariate regression. 
The reduction in $W_{min}(t)$ obtained by increasing the rank from $t = t_0$ to $t = t_1$, 
where $t_0 < t_1$ , is given by
\begin{equation}
 W_{\text{min}}(t_0) - W_{\text{min}}(t_1) = \sum_{j = t_0 +1}^{t_1} \lambda_j.
\end{equation}
This relation depends upon $\mathbf{\Gamma}$ only through the eigenvalues $\{\lambda_j\}$ of
\begin{equation}
 \mathbf{N = \Gamma\Sigma_{YX}\Sigma_{XX}^{-1}\Sigma_{XY}\Gamma}
 \label{RRR}
\end{equation}
However, the value of $t$ and hence, the number and nature of those constraints may not be known prior to statistical analysis.

\section*{Number of factors}

Random Matrix Theory~(RMT)is an important framework to deal with limit distributions on eigenvalues.
Historically, RMT was developed to solve complex problems on nuclear physics, and more recently on quantum chaos~~\cite{Guhr1998}.
During the last decades seminal applications of RMT have arisen in the context of mesoscopic physics, biological microarrays, wireless communication and econophysics~\cite{Forrester1994, Lou2006, Telatar1999, Stanley1999, Bouchaud1999}.
A common ingredient of the cited works is the following result, which here is restated in the language of high dimensional statistics.

Let $X$ be a matrix $p\times n$, where the elements $X_{i,j}$  are  i.i.d. random variables  with distribution $N(0,1)$.
Then, when $p,n\rightarrow \infty$, such that $\frac{n}{p}\rightarrow c \in (0,\infty)$, the spectral density of the Wishart matrix $W = n^{-1} XX^{'}$ converge~(a.s.) to the Marcenko-Pastur law~\cite{Marchenko}
   \begin{equation}
  \rho(x) = \frac{\sqrt{(x_{max}-x)(x-x_{min})}}{2\pi c x},
 \end{equation}
 where 
  \begin{equation}
  x^{max}_{min} = (1\pm \sqrt{c})^2.
  \label{Bound}
  \end{equation}

In the econophysics community, the Marchenko-Pastur distribution is known as a universal result of the Wishart matrices.
If there is no correlation between financial variables then the eigenvalues of its correlation matrix should be bounded between this RMT prediction~\cite{Stanley1999, Bouchaud1999}.

In the field of statistics is of primary importance to consider null hypothesis tests.
The Wishart matrices which appear in the last result can be denoted as $W_p(n,\mathrm{I})$, where $\mathrm{I}$ is the covariance matrix of the population distribution of $n^{-1} XX^{'}$. 
In our case it is of interest to test the hypothesis of identity covariance matrix
$H_0: \Sigma = \mathrm{I}$ against an alternative case  $H_A : \Sigma \neq \mathrm{I}$, where $\Sigma$ has some more general structure.
Under this approach, it is possible to compute a confidence interval to accept or reject the universal result of Wishart matrices of empirical datasets for the general range of dimensions $p$ and $n$.
The approach to quantify a confidence level is based on the approximation  to the null hypothesis distribution of the largest sample eigenvalue $\hat{\lambda}_1$
  \begin{equation}
    P\{\hat{\lambda}_1 > t: H_0 \sim W_p(n,\mathrm{I})\}.
  \end{equation}
The following result of Random matrix theory leads to the needed approximate distribution~\cite{Tracy}.

Assume $A\sim W_p(n, \mathrm{I})$, $p/n \rightarrow \gamma \in (0,\infty)$, and denote $\hat{\lambda}_1$ as the largest eigenvalue in the eigenvalue equation $Au = \hat{\lambda} u$. Then,  the distribution of the largest eigenvalue
approaches to one of the Tracy–Widom $F_{\beta}$ laws
  \begin{equation}
   P\{ n\hat{\lambda}_1 \leq \mu_{np} + \sigma_{np} s | H_0\} \rightarrow F_{\beta}(s)
  \end{equation}
   where $\mu_{np} = (\sqrt{n}+ \sqrt{p})^2$, $\sigma_{np} = \mu_{np}\left( \frac{1}{\sqrt{n}}+\frac{1}{\sqrt{p}} \right)^{1/3}$. 
There exist elegant formulas to solve the Tracy-Widom distribution functions
\begin{equation}
  \begin{split}
   F_1(s) &= \sqrt{F_2(s) \exp\left( -\int_s^{\infty} q(x) dx \right)}\\
   F_2(s) &= \exp\left( -\int_s^\infty (x-s)^2 q(x) dx\right),
  \end{split}
  \label{TracyEqs}
\end{equation}
which are in terms of the solution for $q$ of the non-linear second-order differential equation
$q'' = sq + 2q^3$, $q(s) \sim A_i(s)$ as  $s\rightarrow \infty$, also know as the classical Painlevé type II equations.
The family of functions $F_{\beta}$ are found numerically as a function of $q$.
Despite requiring somewhat effort to solve $F_{\beta}$, from the point of view
of applied data analysis, they are special functions like the normal curve~\cite{Edelman2013}.

Let us exemplify the relevance of the Tracy-Widom test.
Suppose that in a sample of $n = 10$ observations from a $p = 10$ variate Gaussian distribution $N_{10}(0,\Sigma)$, a largest sample eigenvalue $\lambda_1 = 4.25$ emerges.
With these dimensions, the support of the Marchenko-Pastur distribution is bounded into the interval $[0, 4]$~(see Eq.~(\ref{Bound})) 
Then, the question in statistical terms is, an observed largest eigenvalue of $4.25$ is consistent with $H_0: \Sigma = \mathrm{I}$, when $n=p=10$? 
The second order Tracy–Widom approximation~\cite{Johnstone2001} yields a 6\% chance of seeing a value more extreme than 4.25 even if no structure is present, i, e., $\Sigma = \mathrm{I}$.
Against the traditional 5\% benchmark, this is not strong enough evidence to reject the null hypothesis $H_0$~\cite{Johnstone2006}.

The Tracy-Widom test becomes relevant to the determination of the number of components that must be retained in Principal Component Analysis~(PCA), especially,  in the context of high dimensional data, i.e, when $\mathcal{O}(n/p) = 1$.
Beyond PCA, there are several classical problems in multivariate statistics that can take advantage of this type of test.
These problems can be generalized under the greatest root distribution. 
It describes the null hypothesis of apparently different problems, including multiple response linear regression, multivariate analysis of variance, canonical correlations, equality of covariance matrices, among others~\cite{Johnstone2009}.
The next definition from~\cite{Mardia1979} state formally the greatest root distribution.

Let $\mathbf{A} \sim W_p(m,\mathbf{I})$ be independent of $\mathbf{B}\sim W_p(n, \mathbf{I})$, where $m\geq p$. Then the largest eigenvalue $\theta$ of $(\mathbf{A+B})^{-1}\mathbf{B}$ is called the greatest root statistics and is distribution is denoted as $\theta(p,m,n)$.
It has the property
\begin{equation}
 \theta(p,m,n) \overset{d}{=} \theta(n,m+n-p,p),
\end{equation}
useful when $n<p$.

There exist an interesting connection between the greatest root statistics and Tracy-Widom distributions.
In the work of Johnstone~\cite{Johnstone2008} it is shown that with appropriate centering and scaling, the logit transform $W$ of $\theta$ is approximately Tracy–Widom distributed
\begin{equation}
 \frac{W(p,m,n)-\mu(p,m,n)}{\sigma(p,m,n)} \overset{d}{\rightarrow} F_1,
 \label{Johnstone1}
\end{equation}
where
\begin{equation}
 W(p,m,n) = \text{logit}\theta(p,m,n) = \log\left(\frac{\theta(p,m,n)}{1-\theta(p,m,n)}\right)
 \label{Johnstone2}
\end{equation}
is the logit transfor of $\theta$, and the centering and scaling parameters are defined by
\begin{equation}
   \mu(p,m,n) = 2\log\tan\left(\frac{\phi+\gamma}{2}\right),
   \sigma^{3}(p,m,n) = \frac{16}{(m+n-1)^2}\frac{1}{\sin^{2}(\phi+\gamma)\sin\phi\sin\gamma},
   \label{Johnstone3}
\end{equation}
being the angle parameters $\gamma$, $\phi$  defined as
\begin{equation}
  \sin^2\left(\frac{\gamma}{2}\right) = \frac{min(p,n)-1/2}{m+n-1},
  \sin^2\left(\frac{\phi}{2}\right) = \frac{max(p,n)-1/2}{m+n-1}.
  \label{Johnstone4}
\end{equation}

At this point, we are interested to indicate a procedure to determine the $t$ parameter in the RRR model through the greatest root statistics, which conciliate both frameworks.
This commonplace is settled on the Canonical Correlation Analysis~(CCA).
It involves partitioning a collection of variables into two sets.
Let say, a $\mathbf{X}$-set with $q$ variables and a $\mathbf{X}$-set with $p$ variables. 
The purpose is to find maximally correlated combinations $\eta = \mathbf{a'x}$ and $\phi = \mathbf{b'y}$.
Even though CCA has maximal properties similar to PCA, the objective of canonical correlation is on the relationship between two groups of variables instead of interrelationships within a set of variables.

Suppose that ($\mathbf{X,Y}$) is a data matrix of $n$ observations on $q+p$ variables such that each sample is independent of the others and has the populations distribution $N_{p+q}(\mathbf{\mu,\Sigma})$.
Assume the sample covariance matrix $\mathbf{S}$ partitioned
\begin{equation}
   \mathbf{S} = \left(\begin{array}{cc}
   \mathbf{S_{XX}} & \mathbf{S_{XY}} \\
   \mathbf{S_{YX}} & \mathbf{S_{YY}}
   \end{array}\right).
\end{equation}   
The sample squared  canonical correlations ($r_i^2$) for $i=1,\dots,k = \text{min}(p, q)$ are found as the eigenvalues of  $\mathbf{M_S = S_{YY}^{-1}S_{YX}S_{XX}^{-1}S_{XY}}$, 
whereas the population counterpart are given by the eigenvalues of $\mathbf{M_{\Sigma} = \Sigma_{YY}^{-1}\Sigma_{YX}\Sigma_{XX}^{-1}\Sigma_{XY}}$~\cite{Mardia1979}.
Notice that the non-zero eigenvalues of $\mathbf{M_{\Sigma}}$ are the same as the non-zero eigenvalues of $\mathbf{N}$ in Eq.~(\ref{RRR}) for $\mathbf{\Gamma = \Sigma_{YY}^{-1}}$, which is precisely the CCA case in the RRR general model. 

We are now interested in describing the procedure to test the null hypothesis of independence of the two sets of variables $H_0: \mathbf{\Sigma_{12}} = 0$ through the Tracy-Widom test. 
First, let us point out the next result concerning joint independence of partitioned Wishart matrices.

Let $\mathbf{M}\sim  W_{p}(n, \mathbf{\Sigma})$, and partition the matrix $\mathbf{M}$ into the submatrices $\mathbf{M}_{11}$ of dimensions $a\times a$ and $\mathbf{M}_{22}$ of dimensions $b\times b$, where $a+b = p$ and $n>a$. 
Define the product of matrices $\mathbf{M}_3 = \mathbf{M}_{22} - \mathbf{M}_{21}\mathbf{M}_{11}^{-1}\mathbf{M}_{12}$. Then~\cite{Mardia1979}

\begin{itemize}
 \item[(a)] $\mathbf{M}_3$ has the $W_b(n-a,\mathbf{\Sigma_3})$ distribution and is independent of $(\mathbf{M}_{11},\mathbf{M}_{22})$, 
 \item[(b)] if $\mathbf{\Sigma}_{12} = 0$, then $\mathbf{M}_{22} - \mathbf{M}_{3} = \mathbf{M}_{21}\mathbf{M}_{11}^{-1}\mathbf{M}_{12}$ has the $W_b(a,\mathbf{\Sigma}_{22})$ distribution, and $\mathbf{M}_{21}\mathbf{M}_{11}^{-1}\mathbf{M}_{12}$, $\mathbf{M}_{11}$, and $ \mathbf{M}_{3}$ are jointly independent.
\end{itemize}

On the other hand, the hypothesis technique of Union Intersection Test~(UIT) uses the statistics based on the largest eigenvalue $r_1^2$ of $\mathbf{M_S}$.

But $\mathbf{M_S}$ can be written as $[\mathbf{M}_{3} + (\mathbf{M}_{22}-\mathbf{M_{3}})]^{-1}(\mathbf{M}_{22}-\mathbf{M}_{3})$, where $\mathbf{M}_{22}= n\mathbf{S}_{YY}$, $\mathbf{M}_3 = n(\mathbf{S_{YY}-S_{YX}S_{XX}^{-1}S_{XY}})$, and  $\mathbf{M}_{22}-\mathbf{M}_{3}$ satisfies the independence condition of the greatest root statistics.
Therefore, under $H_0: \mathbf{\Sigma_{12}} = 0$, $r_1^2$ has the $\theta(p, n  - q  - 1, q)$ distribution, and the Tracy–Widom approximation can be applied.

The previous derivation shows a procedure to statistically determine the rank of a RRR model through the framework of RMT. 
Specifically, it has been delineated the connection of the Tracy-Widom distribution to test the null hypothesis $H_0: \mathbf{\Sigma_{12}} = 0$ in the particular case CCA of the general RRR models.
In what follows, it is described the applied methodology to find the number of significative components or factors in the CCA using our data sets of predictor and response cryptocurrencies variables.

The first step to use these techniques in real data is based on numerically solving the system of equations involved in Eq.~(\ref{TracyEqs}) taking into account the Painlevé equations with the boundary condition that as $t\rightarrow\infty$, $q(t)$ is asymptotic to the Airy function $Ai(t)$. 
We solve these non-linear differential equations with an absolute tolerance error of $1\times 10^{-15}$ following the  approach given in~\cite{Edelman2013}.  
In Table~\ref{table3} it is shown  a subsample of the kind of values obtained. 
The first and second columns display the x,y values on the plane of the Tracy-Widom distribution, respectively. 
The third column shown the cumulative density value~(cdv) corresponding to the related $x$,$y$ values, which subtracted from 1 determine the level of significance in the statistical test of Tracy-Widom.  
\begin{table}[!ht]
\centering
\caption{
{\bf Tracy-Widom values }}
\begin{tabular}{|l|l|l|}
\hline
\textbf{x} & \textbf{y} & \textbf{cdv} \\ \hline
\vdots     & \vdots  & \vdots \\ \hline
1.995 & 0.017669 & 0.989510 \\ \hline
2.000 & 0.017535 & 0.989598 \\ \hline
2.005 &	0.017402 & 0.989685 \\ \hline
2.010 & 0.017270 & 0.989771 \\ \hline
2.015 &	0.017139 & 0.989857 \\ \hline
2.020 &	0.017009 & 0.989942 \\ \hline
2.025 &	0.016880 & 0.990026 \\ \hline
2.030 &	0.016751 & 0.990110 \\ \hline
2.035 &	0.016623 & 0.990193 \\ \hline
2.040 &	0.016497 & 0.990276 \\ \hline
\vdots & \vdots & \vdots    \\ \hline
\end{tabular}
\begin{flushleft} 
Subsample of the $x$,$y$ values on the plane, and the corresponding cdv of the Tracy-Widom distribution.
\end{flushleft}
\label{table3}
\end{table}

\begin{figure}[!h]
\includegraphics[height=0.3\textheight]{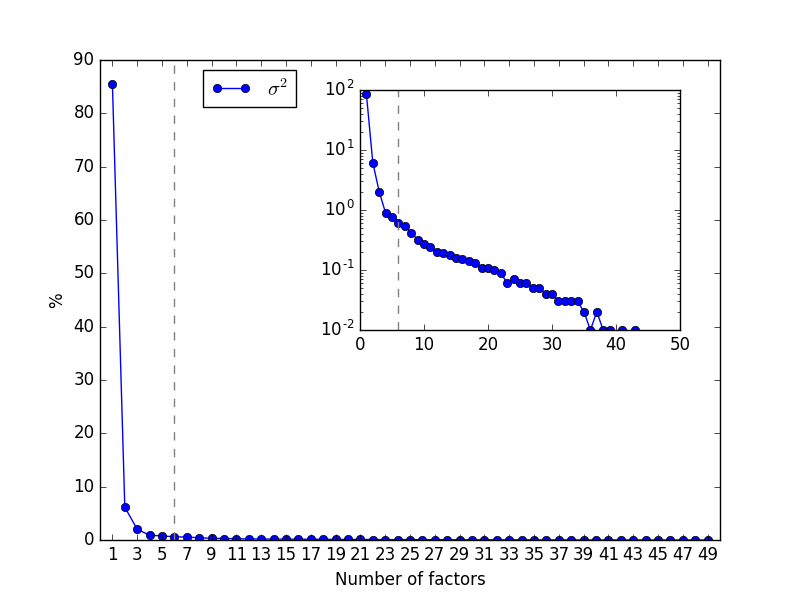}
\caption{{\bf Explained variance in CCA.}
Variance contribution in percentage  a function of the component element. The inset graph show the same but in semi-log scale.}
\label{fig4}
\end{figure}

Next, we apply  CCA to the set of cryptocurrencies variables. 
In this analysis predictor and responses variables  previously found in the variable selection section were considered as the $\mathbf{X,Y}$ sets, respectively.
When using the greatest root distribution $\theta(p, n  - q  - 1, q)$ with parameters $p=49$, $q=51$, and $n=4532$ trough Eqs.~(\ref{Johnstone1}-\ref{Johnstone4}) 
it is found $6$ factors at the significance level of  $0.01$.
In Fig~\ref{fig4} it is shown the explained variance in percentage as a function of the number of factors, which in CCA case the increment on the predictor and response components it is considered symmetrically, but the fixed lag time of $\Delta t = 1$ provides the forecasting element.
There, the dashed vertical gray line represents the cut where it is found the number of significant components.
The inset graph shows the same but in a semi-log scale.
The plot does not show an abrupt change in the curve.
Thereby, if we use the elbow criterion, would not be possible to determine the appropriate number of components to consider in the model. 
Moreover, comparing with the cross-validations approach, the computational time of the Tracy-Widom test is negligible, since we only need to compute once the table of significance level.

\begin{figure}[!h]
\includegraphics[height=0.3\textheight]{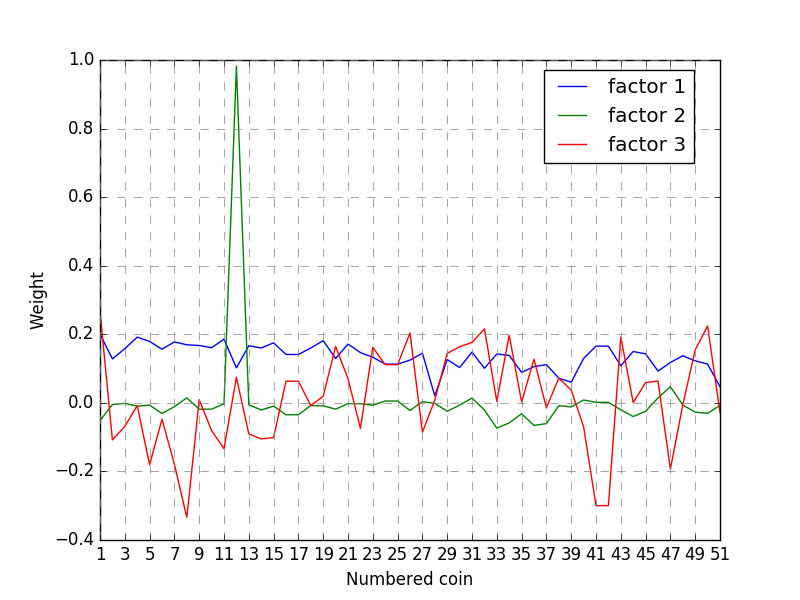}    
\caption{{\bf Response weights.}
Eigenvector components associated to the response variables.}
\label{fig5}
\end{figure}

\begin{figure}[!h]
\includegraphics[height=0.3\textheight]{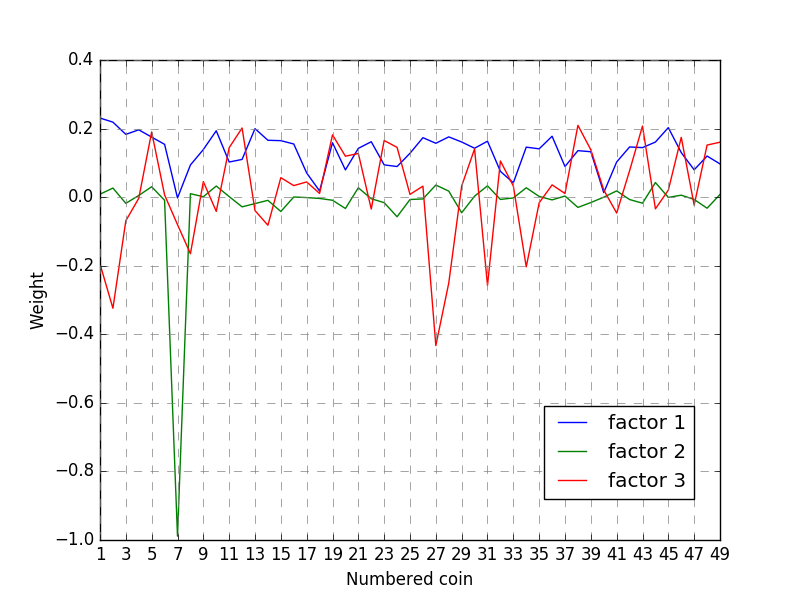}
\caption{{\bf Predictor weights.}
Eigenvector components associated to the predictor variables.}
\label{fig6}
\end{figure}

Furthermore, we plot the response and predictor weights of the first three factors in Figs~\ref{fig5} and \ref{fig6}.
It can be seen that all the coefficients of the first factor have positive weights in both response and predictor cases.
It resembles the behavior of financial indices under PCA, where the eigenvector~(or factor) associated to the largest eigenvalue only has positive coefficients, and is named the collective mode.
Inspired in this logic and since the first pair of response-predictor factors are associated to the largest singular value, we can label them as the collective-response and collective-predictor modes, respectively.
The second pair of factors, shows as green in the same figures, have different behavior.
In general, they fluctuate around zero but have a strong peak in a specific currency.
In the response case, this peak is positive and correspond to the \emph{vechain} coin, whereas for the predictor case the peak is negative and correspond to the \emph{tether} coin.
Based on these results, we could venture to call these factors the specific-response and specific-predictor modes, but it is necessary more evidence from a dynamic analysis to hold this observation.
Finally, the third pair of response-predictor factors do not show a specific pattern and is not possible to try to give them a meaning. The following fourth to sixth factors presented similar behavior and was the reason why we omit them in the Figs~\ref{fig5} and \ref{fig6}.

A usual question about the determination of the number of factors in the scientific community nonfamiliar with econometric problems is about why not use as most as possible factors or components since this could increment the precision of the forecast.
Thereupon, it is worth to make a comment in this direction. 
In econometrics, it is fundamental to determine the minimum number of components in a model because it is wanted to attribute explanatory meaning to each component in order to explain the economic theory behind them. Therefore, the concerns of the proper determination in the number of components in this study.

\section*{Conclusion}

In general, random matrices seems to be a promising tool to deal with factor determinations in financial and economic problems.
Nevertheless, much theory has been developed around random matrices which is still not applied by the practitioners.
With the intention to fill this gap, we described the connection between the RRR models and the Tracy-Widom test to determine the number of significative factors or components in the reduced CCA case of the general RRR models. The results show an interpretable meaning for the first two pairs of response-predictor set of cryptocurrencies variables. 
The main advantage of the proposed procedure is to avoid the subjective element of visual inspection as is the elbow criterion, and abstain from the computational cost of the cross-validation approach. 
Beyond this, the distributional test of Tracy-Widom has the conceptual advantage of its relationship with a more general mathematical framework, which touches many branches of fundamental mathematics and theoretical physics.

Another contribution of this work is the variables selection methodology based on information theory.
We use TE to measure the flow of information between cryptocurrencies return variables.
Since TE can be seen as a generalization of  Granger causality test under some circumstances, we can cover a lot of scenarios including possible non-linear dependencies between the variables.
We propose a heuristic criterion related to the in-degree and out-degree of the nodes when the TE estimation is seeing as a graph. 
Again, the symbol approach to measure TE has the advantage of having a distributional test.
Therefore, our selected set of response and predictor variables have associated a \textit{p-value}, which is always desired in the econometric community, and make our results more robust in the statistical sense.

Interesting future work is to consider the case when $\mathbf{\Sigma \neq I}$ in order to model heteroscedasticity and serial correlations for example. 
This kind of structure can be modeled using free matrices to obtain the factors as an optimization problem.
Also, it can be solved by numerical simulations, where the Tracy-Widom joint distribution of eigenvalues  plays an crucial role. 
Such problems are related to the well know dynamics factor models in the econometrics literature, and have the advantage to be more explanatory and linked with structural forecast models like Vector Autoregressive~(VAR) and Vector Error Correction Model~(VECM).

\section*{Acknowledgments}
Funding from CONACYT through fund FOSEC SEP-INVESTIGACIÓN BÁSICA 252996  is acknowledged. 

\nolinenumbers

\section*{Supporting information}

\paragraph*{S1 File.}
\label{S1_File}
{\bf Raw data.} Prices of cryptocurrencies used in this study as described in section Data before preprocessing.

\paragraph*{S1 Table.}
\label{S1_Table}
{\bf List names of cryptocurrencies.}  Listed from highest to lowest capitalization as it was at February 2018.
\begin{table}[!ht]
\centering
\begin{tabular}{|l|l|l|l|}
\hline
\textbf{number} & \textbf{name} & \textbf{number} & \textbf{name}\\ \hline
1  & bitcoin          & 51  & revain                \\
2  & ethereum         & 52  & electroneum           \\
3  & ripple           & 53  & digixdao              \\
4  & Bitcoin cash     & 54  & gas                   \\
5  & litecoin         & 55  & byteball              \\
6  & cardano          & 56  & Basic attention token \\
7  & neo              & 57  & dragonchain           \\
8  & stellar          & 58  & digibyte              \\
9  & eos              & 59  & loopring              \\
10 & iota             & 60  & Golem network tokens  \\
11 & dash             & 61  & zilliqa               \\
12 & nem              & 62  & bytom                 \\
13 & monero           & 63  & Kyber network         \\
14 & lisk             & 64  & monacoin              \\
15 & Ethereum classic & 65  & pivx                  \\
16 & tron             & 66  & syscoin               \\
17 & vechain          & 67  & aelf                  \\
18 & qtum             & 68  & dentacoin             \\
19 & Bitcoin gold     & 69  & qash                  \\
20 & tether           & 70  & bitcore               \\
21 & icon             & 71  & cryptonex             \\
22 & omisego          & 72  & nebulas\_token        \\
23 & zcash            & 73  & ethos                 \\
24 & raiblocks        & 74  & pillar                \\
25 & Binance coin     & 75  & Power ledger          \\
26 & steem            & 76  & iostoken              \\
27 & populous         & 77  & gxshares              \\
28 & verge            & 78  & factom                \\
29 & Bytecoin bcn     & 79  & aion                  \\
30 & stratis          & 80  & salt                  \\
31 & siacoin          & 81  & dent                  \\
32 & rchain           & 82  & funfair               \\
33 & dogecoin         & 83  & kin                   \\
34 & status           & 84  & nxt                   \\
35 & waves            & 85  & cindicator            \\
36 & bitshares        & 86  & zcoin                 \\
37 & maker            & 87  & Enigma project        \\
38 & walton           & 88  & neblio                \\
39 & 0x               & 89  & Polymath network      \\
40 & decred           & 90  & wax                   \\
41 & aeternity        & 91  & chainlink             \\
42 & augur            & 92  & reddcoin              \\
43 & komodo           & 93  & maidsafecoin          \\
44 & veritaseum       & 94  & Request network       \\
45 & hshare           & 95  & bancor                \\
46 & ucash            & 96  & tenx                  \\
47 & kucoin\_shares   & 97  & smartcash             \\
48 & ardor            & 98  & santiment             \\
49 & zclassic         & 99  & particl               \\
50 & ark              & 100 & blocknet             \\
\hline
\end{tabular}
\begin{flushleft} 
\end{flushleft}
\end{table}

\paragraph*{S2 Table.}
\label{S2_Table}
{\bf Entire list of predictor-response variables.} Each set is ordered from highest to lowest capitalization.
\begin{table}[!ht]
\centering
\begin{tabular}{|l|l|l|l|}
\hline
\textbf{number} & \textbf{predictor} & \textbf{number} & \textbf{response}\\ \hline
1  & ethereum             & 1  & bitcoin               \\
2  & neo                  & 2  & ripple                \\
3  & dash                 & 3  & Bitcoin cash          \\
4  & monero               & 4  & litecoin              \\
5  & lisk                 & 5  & cardano               \\
6  & Bitcoin gold         & 6  & stellar               \\
7  & tether               & 7  & eos                   \\
8  & steem                & 8  & iota                  \\
9  & populous             & 9  & nem                   \\
10 & siacoin              & 10 & Ethereum classic      \\
11 & rchain               & 11 & tron                  \\
12 & dogecoin             & 12 & vechain               \\
13 & bitshares            & 13 & qtum                  \\
14 & 0x                   & 14 & icon                  \\
15 & augur                & 15 & omisego               \\
16 & komodo               & 16 & zcash                 \\
17 & veritaseum           & 17 & raiblocks             \\
18 & ucash                & 18 & Binance coin          \\
19 & Kucoin shares        & 19 & verge                 \\
20 & revain               & 20 & Bytecoin bcn          \\
21 & digixdao             & 21 & stratis               \\
22 & gas                  & 22 & status                \\
23 & byteball             & 23 & waves                 \\
24 & dragonchain          & 24 & maker                 \\
25 & loopring             & 25 & walton                \\
26 & Golem network tokens & 26 & decred                \\
27 & zilliqa              & 27 & aeternity             \\
28 & bytom                & 28 & hshare                \\
29 & Kyber network        & 29 & ardor                 \\
30 & pivx                 & 30 & zclassic              \\
31 & aelf                 & 31 & ark                   \\
32 & dentacoin            & 32 & electroneum           \\
33 & cryptonex            & 33 & Basic attention token \\
34 & Nebulas token        & 34 & digibyte              \\
35 & ethos                & 35 & monacoin              \\
36 & funfair              & 36 & syscoin               \\
37 & kin                  & 37 & qash                  \\
38 & nxt                  & 38 & bitcore               \\
39 & zcoin                & 39 & pillar                \\
40 & Enigma project       & 40 & Power ledger          \\
41 & neblio               & 41 & iostoken              \\
42 & chainlink            & 42 & gxshares              \\
43 & maidsafecoin         & 43 & factom                \\
44 & Request network      & 44 & aion                  \\
45 & bancor               & 45 & salt                  \\
46 & tenx                 & 46 & dent                  \\
47 & santiment            & 47 & cindicator            \\
48 & particl              & 48 & Polymath network      \\
49 & blocknet             & 49 & wax                   \\
   &                      & 50 & reddcoin              \\
   &                      & 51 & smartcash            \\
\hline   
\end{tabular}
\begin{flushleft} 
\end{flushleft}
\end{table}

\end{document}